\documentstyle[aps,12pt,psfig]{revtex}

\newcommand{ \be }{\begin{equation}}
\newcommand{ \ee }{\end{equation}}
\newcommand{ \bea }{\begin{eqnarray}}
\newcommand{ \eea }{\end{eqnarray}}
\newcommand{ \la }{\langle}
\newcommand{ \ra }{\rangle}
\newcommand{ \mpx }{\langle p_{x} \rangle}

\begin{document}
\normalsize

\title{ Proton and Pion Production Relative to the Reaction Plane
in Au + Au Collisions at AGS Energies}

\author{
  J.~Barrette$^5$, R.~Bellwied$^{9}$, 
  S.~Bennett$^{9}$, R.~Bersch$^7$, P.~Braun-Munzinger$^2$, 
  W.~C.~Chang$^7$, W.~E.~Cleland$^6$, M.~Clemen$^6$, 
  J.~Cole$^4$, T.~M.~Cormier$^{9}$, 
  Y.~Dai$^5$, G.~David$^1$, J.~Dee$^7$, O.~Dietzsch$^8$, M.~Drigert$^4$,
  K.~Filimonov$^5$, S.~C.~Johnson$^7$, 
  J.~R.~Hall$^9$, T.~K.~Hemmick$^7$, N.~Herrmann$^2$, B.~Hong$^2$, 
  Y.~Kwon$^7$,
  R.~Lacasse$^5$, Q.~Li$^{9}$, T.~W.~Ludlam$^1$,
  S.~K.~Mark$^5$, R.~Matheus$^{9}$, S.~McCorkle$^1$, J.~T.~Murgatroyd$^9$,
  D.~Mi\'{s}kowiec$^2$,
  E.~O'Brien$^1$,  
  S.~Panitkin$^7$, T.~Piazza$^7$, M.~Pollack$^7$, C.~Pruneau$^9$, 
  M.~N.~Rao$^7$, E.~Reber$^4$, M.~Rosati$^5$, 
  N.~C.~daSilva$^8$, S.~Sedykh$^7$, U.~Sonnadara$^6$, J.~Stachel$^3$, 
  E.~M.~Takagui$^8$, 
  S.~Voloshin$^3\footnote{On leave from Moscow Engineering Physics Institute,
     Moscow, 115409,  Russia}$,
  T.~B.~Vongpaseuth$^7$,
  G.~Wang$^5$, J.~P.~Wessels$^3$, C.~L.~Woody$^1$, 
  N.~Xu$^7$,
  Y.~Zhang$^7$, C.~Zou$^7$\\ 
(E877 Collaboration)
}
\address{
 $^1$ Brookhaven National Laboratory, Upton, NY 11973\\
 $^2$ Gesellschaft f\"ur Schwerionenforschung, 64291 Darmstadt, Germany\\
 $^3$ Universit\"at Heidelberg, 69120 Heidelberg, Germany\\
 $^4$ Idaho National Engineering Laboratory, Idaho Falls, ID 83402\\
 $^5$ McGill University, Montreal, Canada\\
 $^6$ University of Pittsburgh, Pittsburgh, PA 15260\\
 $^7$ SUNY, Stony Brook, NY 11794\\
 $^8$ University of S\~ao Paulo, Brazil\\
 $^9$ Wayne State University, Detroit, MI 48202\\
}

\date{\today}

\maketitle

\begin{abstract}
Results are presented of an analysis of proton and charged pion 
azimuthal distributions measured with respect to the reaction plane 
in Au + Au collisions at a beam momentum of about 11$A$ GeV/c.
The azimuthal anisotropy is studied as a function of particle rapidity 
and transverse momentum for different centralities of the collisions.
The triple differential (in rapidity, transverse momentum, and
azimuthal angle) distributions are reconstructed.
A comparison of the results with a previous analysis of charged particle
and transverse energy flow as well as with model predictions are presented. 

\end{abstract}
\narrowtext


\section{Introduction}

Collective phenomena play an important role in heavy-ion collisions, but
for a long time it was assumed that, at collision energies much greater
than 1~GeV/nucleon, only longitudinal and azimuthally symmetric
transverse radial flow would survive.  During the last few years the
situation has changed qualitatively.  Anisotropic, directed as well as
elliptic, flow has been observed at the BNL
AGS~\cite{l877flow1,l877flow2}, and strong indications of elliptic flow
at the CERN SPS~\cite{lna49} have been demonstrated.  The theoretical
understanding of the effect and model calculations involving anisotropic
flow have progressed significantly; different anisotropic flow patterns
have been linked to such phenomena as quark-gluon plasma formation,
softening of the equation of state~\cite{lbrav,lshur,lko1,lris}, and
mean-field effects during the fireball
evolution~\cite{lko2,lsorge1,lrqmd23}.  The appropriate tools for flow
studies at high energies have been developed~\cite{lolli,lvzh}.  It was
also noticed that anisotropic flow could be important for other
measurements, such as two-particle correlations~\cite{lmisko,lvc}.
Anisotropic flow has become an essential part of the global picture of
heavy-ion collisions at ultrarelativistic energies~\cite{lstachel}; it
is considered one of the key tools to elucidate the dynamics of the
collision.

In the current paper we present results of the analysis of anisotropic
transverse collective flow of identified particles, protons and charged
pions, detected in the E877 spectrometer in Au + Au collisions at a beam
momentum of 10.8 and 11.4$A$ GeV/c.  The data were taken during the 1993
and 1994 AGS heavy-ion runs.  Using calorimeter data we reconstruct the
reaction plane event-by-event, and analyze the particle production with
respect to this reaction plane.  A similar analysis of charged particle
multiplicity and transverse energy flow presented in~\cite{l877flow2}
displayed a strong directed flow as well as an elliptic flow with the
primary axis in the reaction plane, {\em i.e.}  perpendicular to the
direction of the ``squeeze-out'' effect observed at lower
energies~\cite{squeeze}.  Here, using the E877 spectrometer data, we
apply the same procedure to determine the triple differential
distributions $d^{3}N/(dy\,p_t dp_t d\phi )$ of identified protons and
charged pions.

The paper is organized as follows.  After a description of the apparatus
we discuss the procedure of the analysis.  In the subsequent sections we
present first the results of the Fourier analysis of azimuthal particle
distributions in different rapidity bins and for different centralities
of the collision.  These results are, to a large extent, independent of
the uncertainties in the spectrometer efficiency and the detector
acceptance.  Then the results on azimuthal anisotropies and related
quantities (such as the mean transverse momentum projected into the
reaction plane $\la p_x \ra$) are discussed.  Different flow scenarios
are discussed from the point of view of the observed dependence of the
flow signal on the transverse momentum $p_{t}$ of the particle.  We
compare our measurements with the evaluation of nucleon and pion flow
derived from the measurements of charged particle and transverse energy
flow~\cite{l877flow2}, as well as with model predictions (RQMD versions
1.08 and 2.3~\cite{lrqmd23,lrqmd}).

\section{Apparatus}

The E877 apparatus is shown in Fig.~1.  In the E877 setup, charged
particles, emitted in the forward direction and traversing the
collimator ($ -134 < \theta_{horizontal} < 16 $ mrad, $ -11 <
\theta_{vertical} < 11 $ mrad), are analyzed by a high resolution
magnetic spectrometer.  The spectrometer identifies particles via
simultaneous measurements of momentum and velocity.  The momentum of
each particle is measured using two drift chambers (DC2 and DC3,
position resolution about 300~$\mu$m) whose pattern recognition is aided
by four multi-wire proportional chambers (MWPC).  The average momentum
resolution is $\Delta p/p \approx$3\% limited by multiple scattering.  A
time-of-flight hodoscope (TOFU) located behind the tracking chambers
provides the time-of-flight with a typical resolution of 85~ps.  The
spectrometer acceptance covers mostly the forward rapidity region with
transverse momentum coverage including $p_t=0$.  Further discussion of
the acceptance for different particle species can be found
in~\cite{l877pid}.

A clean particle identification is particularly crucial for the study of
flow since the extracted signal may be different for different particle
species both in magnitude and in sign.  A small admixture in the
particle sample of other particle species could therefore heavily bias
the final results.  Consequently, in the current analysis, strict
requirements are applied for particle identification.  Using the
measured momentum and velocity of the particle we calculate the particle
mass.  In a plot of momentum {\it vs.} the square of the particle mass
$m^2$ we select entries in the region of $\pm 1.5\,\sigma_{m^2}(p)$
around the $m^2$ peak of the particle under study and reject all entries
within a $3\,\sigma_{m^2}(p)$ region of another particle species. Here,
$\sigma_{m^2}(p)$ represents the standard deviation of the (Gaussian)
mass-squared distribution at a given particle momentum $p$.

The determination of the centrality of the collision and of the reaction
plane orientation are made using the transverse energy flow measured in
the target calorimeter (TCal), and participant calorimeter (PCal).  Both
calorimeters have $2 \pi$ azimuthal coverage and, combined together,
they provide nearly complete polar angle coverage as viewed from the
nucleus-nucleus center of mass system: TCal and PCal cover the
pseudorapidity regions $-0.5 <\eta <0.8$ and $0.8 <\eta < 4.2$,
respectively~\cite{l877flow2,l877et}. The pseudo-rapidity $\eta$ = - ln
tan($\theta/2$) is defined in terms of the polar angle $\theta$ in the
laboratory frame.

\section{Analysis}

The structure of the analysis is very similar to that used in
\cite{l877flow2}, and many details can be found there.  Here, the
distributions $d^{3}N/(p_t dp_t dy d\phi )$ of identified protons and
charged pions are generated in a coordinate system where the $x$- and
$z$-axes span the reaction plane.  The azimuthal angle in this system is
defined as $\phi=\phi_{lab}-\Psi_1$, where $\Psi_1$ is the reaction
plane angle, measured for every event using the direction of the
transverse energy flow in TCal and PCal, and $\phi_{lab}$ is the
azimuthal angle of an individual particle in the laboratory frame.  The
$x$-axis is defined in such a way that it points in the direction of the
transverse energy flow at forward rapidities.  The transverse momentum
components $p_{x}$ and $p_{y}$ of identified particles are evaluated in
this coordinate system, and the particle distributions with respect to
the reaction plane are constructed.  The azimuthal anisotropy of
particle production is studied by means of Fourier analysis of azimuthal
distributions~\cite{l877flow1,l877flow2,lvzh}.  This yields the
rapidity, transverse momentum, and centrality dependence of the Fourier
coefficients $v_n$ (amplitude of $n$-th harmonic) in the decomposition:
\be
E \frac{d^3 N}{d^3 p} = 
\frac{d^3N}{p_t d p_t dy d\phi}=
\frac{1}{2\pi} \frac {d^2N}{p_t d p_t dy}(1+2 v_1' \cos (\phi)
+2 v_2' \cos(2 \phi)+2 v_3' \cos(3 \phi)+...).
\label{ed3}
\ee
The E877 spectrometer provides full $2\pi$ acceptance in the spectrometer
only for a very limited $p_{t}$ range (approximately $p_t \le 50$~MeV/c), 
but the triple differential multiplicity is constructed in full, using
the $2\pi$ range of the reaction plane angle distribution.
Note that the coverage of the calorimeters used for the reaction 
plane determination does not overlap with the spectrometer coverage, 
and thus the analysis is largely free from problems 
related to auto-correlations.

Similarly to the analysis presented in~\cite{l877flow2}, the reaction
plane angle $\Psi_1$ is determined in four non-overlapping
pseudorapidity windows. The 'reaction plane resolution', i.e. the
accuracy with which the reaction plane orientation is determined, is
evaluated by studying the correlation between flow angles determined in
different windows.  Finally, the flow signals are corrected for this
resolution.  Details of this procedure are described
in~\cite{l877flow2}.  In short, the true value of each Fourier
coefficient $v_n$ is obtained by:
\be
v_n=v_n'/\la \cos(n(\Psi_1-\Psi_R) \ra,
\ee 
where $v_n'$ is the observed signal, and $\la \cos(n(\Psi_1-\Psi_R) \ra$
is the mean cosine of the difference of the reconstructed ($\Psi_1$) and
true ($\Psi_R$) reaction plane angles, characterizing the reaction plane
resolution. The values of $\la \cos(n(\Psi_1-\Psi_R) \ra$ are evaluated
as outlined in \cite{l877flow2}. As a result we obtain the true
azimuthal distribution with respect to the reaction plane orientation:
\be
E \frac{d^3 N}{d^3 p} = 
\frac{1}{2\pi} \frac {d^2N}{p_t d p_t dy}(1+2 v_1 \cos(\phi_{lab}-\Psi_R)
+2 v_2 \cos(2(\phi_{lab}-\Psi_R))+...).
\label{ed3t}
\ee

The E877 spectrometer has a relatively small azimuthal coverage.  Due to
this, the analysis of azimuthal anisotropies is rather sensitive to
biases of different kinds, which could simulate an event anisotropy and
propagate to the final results.  For instance, during the off-line
analysis, it was found that a small fraction of the recorded events
($\leq$~2--4\%) have an anomalously high number of hits in the drift
chambers.  In these events the hit density was too high to perform
reliable tracking and the high occupancy of the spectrometer was found
to be correlated with the orientation of the reaction plane.  A bias due
to this was avoided by removing these events from the analysis
completely and not only from the sample used to generate spectra.  The
tracking efficiency in the spectrometer (typically on the order of 90\%)
depends slightly on the spectrometer occupancy.  A special correction
for this effect was developed and checked by a Monte-Carlo simulation.
The correction is based on the weighting of each track in accordance
with the local track densities in the drift chambers and time-of-flight
wall.  Due to different gain factors and dead towers in the calorimeters
the distribution in the reconstructed reaction plane angle is generally not
flat.  Special precaution was taken to make the reaction plane angle
distribution as flat as possible and to remove possible biases (see
Appendix~A).

The Fourier coefficients of azimuthal distributions evaluated by the
procedure described above (and corrected for the reaction plane
resolution), are combined with the measurements of $p_t$ and rapidity
spectra~\cite{l877pid}.  In this way the triple differential
distributions in $y$, $p_t$, and $\phi$ (see Equation~(\ref{ed3t})) are
determined and analyzed.

\section{Results}
\label{sresu}

\subsection*{Directed Flow}

Anisotropic flow reveals itself already in the dependence 
on the azimuthal emission angle of 
the inverse slope parameter of the invariant spectra
$E d^3N/d^3p=d^3N/(p_t dp_t dy d\phi )=d^3N/(m_t dm_t dy d\phi)$
when plotted as a function of $m_t-m$ for a given $\phi$. Here,
$m_t=\sqrt{p_t^2+m^2}$ is the transverse mass.  
We extract the inverse slope parameter $T_B(\phi)$ 
by fitting the spectra in the region $m_t-m > 0.1$~GeV/c$^2$  by 
a thermal (Boltzmann) function 
\be
E \frac{d^3N}{d^3p} \propto   
m_t  \exp \left(- \frac{m_{t}-m}{T_{B}(\phi)}\right).
\label{efun3}
\ee
The shape of the spectra is not perfectly reproduced by this function,
and in order to obtain a good quality description of the spectra in the
entire $m_t$ region, the weights of all points were chosen to be equal
for the fit and not in accordance with statistical errors.

In Fig.~\ref{fspectra} we show the $m_t$ spectra of protons emitted into
the rapidity interval $2.8<y<2.9$ at different angles relative to the
reaction plane together with these thermal fits for different
centralities.  Clearly, the spectra of protons emitted in the direction
of flow ($\phi$=0) are significantly flatter than those of protons
emitted in the opposite direction ($\phi=\pi$).  To visualize the
angular dependence of the inverse slope parameters the results from the
fit are shown in Fig.~\ref{fteff} for three rapidity intervals. Here
$T_{B}$ is plotted as a function of the azimuthal emission angle.  The
results are presented for four centrality regions, selected in
accordance with PCal $E_T$ and corresponding to the values of
$\sigma_{top}(E_T)/\sigma_{geo} \approx$ 23--13\%, 13--9\%, 9--4\%, and
$<$4\% (see Fig.~4 in~\cite{l877flow2}). The value of
$\sigma_{top}(E_T)$ is obtained by an integration of $d\sigma /dE_T$
from a given value of $E_T$ to the maximal one observed, and the
geometric cross section is defined as $\sigma_{geo}=\pi
r_0^2(A^{1/3}+A^{1/3})^2=6.13$~b, with $A=197$ and $r_0$=1.2~fm.

The results shown in Figs.~\ref{fspectra} and~\ref{fteff} are not
corrected for the reaction plane resolution.  The correction would
increase the difference between the maximal and minimal values of
$T_{B}$ by about a factor of 1.5.  Such a correction (in terms of $T_B$)
is rather complicated; for the quantitative description of the flow
effects we use a Fourier analysis of the azimuthal
distributions~\cite{l877flow1,l877flow2,lvzh} where all corrections are
implemented.

We quantify directed flow by $v_1$, the amplitude of the first harmonic
in the Fourier decomposition of the azimuthal particle distribution
defined in Equation~(\ref{ed3}) and corrected for the reaction plane
resolution.  The coefficient $v_1$ is analyzed as a function of
transverse momentum for different rapidity bins and collision
centralities.  The results for $v_1(p_t)$, corrected for the reaction
plane resolution, are presented in Figs.~\ref{fv1prot}--\ref{fv1pm}
for protons and charged pions.

The error-bars shown in all figures represent statistical errors only. 
The systematic uncertainties have mostly two sources:
i) The uncertainty in the determination of the reaction plane
resolution leading to a relative error in $v_1$ of the order of
5--10\%, similar for all particle species. 
ii) The uncertainty in the occupancy correction, the accuracy of which
we estimate, by inspection of Monte-Carlo simulations, to be of the
order of 20--30\%. The correction itself is different for different
particle species.  It is negligible for positive pions (and for negative
pions from the 1993 run), which are registered in the low occupancy
region of the spectrometer.  The correction is maximal, and reaches
absolute values of about 0.1 for the data shown, for protons at low
$p_t$ and/or high rapidities.  The rapidly increasing uncertainty in the
occupancy correction in the spectrometer region close to the beam limits
our measurements of proton flow at very low $p_t$.  Multiplying the
uncertainty in the occupancy correction with its absolute value we end
up with a systematic error in the absolute value of $v_1$ of about 0.03
where the correction is maximal.
          
As can be seen from Fig.~\ref{fv1prot}, proton emission is very strongly
correlated with the orientation of the reaction plane.  Protons of
larger $p_{t}$ have larger values of $v_{1}$ with some tendency to
saturation in the highest $p_{t}$ region.  The largest flow signal
observed corresponds to a difference in the high $p_t$ proton yields
along ($\phi=0$) and opposite to the flow direction ($\phi=\pi$) of
almost a factor of 10 (as can also be seen in Fig.~\ref{fspectra}).
Both positive (Fig.~\ref{fv1pp}) and negative (Fig.~\ref{fv1pm}) pions
exhibit weak flow in the direction opposite to that of protons (negative
values of $v_1$) over most of the $p_t$ region studied.  The maximum
negative values of $v_1^{\pi}$ are about $-0.1$, significantly less in
magnitude than the flow signal observed for protons.
\footnote{Due to a ``hole'' in the $p_t$ acceptance of negative pions
for the magnetic field polarity used during the 1994 run we have
combined data from the 1994 run and from the lower statistics 1993 run (shown
in Fig.~\ref{fv1pm} as open points), where data exist for both
polarities of the magnetic field.}

Independent of any flow scenario $v_{1}(p_t)$ must vanish at $p_{t}=0$,
due to the continuity of the spectra.  The small non-zero values of
$v_1$ for the lowest $p_t$ bin in Figs.~\ref{fv1pp} and~\ref{fv1pm}
(for low centralities and rapidities close to beam rapidity; positive
for negative pions and negative for positive pions) are mostly due to
the finite bin size of the data.  The details of the behavior of $v_1$
at very low $p_t$ are presented in Fig.~\ref{fpipm} where we show, in
the same plot, the results for positive and negative pions in the
rapidity region $3.2<y<3.6$ for two different centralities and with much
finer binning.

Not only are the sign and magnitude of the flow signal very different
for protons and pions but also the functional dependence on $p_t$ varies
with particle species.  The $v_1$ values for protons grow almost
linearly with $p_t$ over the entire $p_t$ region.  Inspecting the
transverse momentum dependence of $v_1$ for pions one can distinguish
three regions in $p_t$: i) The very low $p_t$ region (below $p_t\approx
0.1$~GeV/c) where in fact the flow signals of positive and negative
pions are different. Positive pions show $v_1$ values decreasing rapidly
and monotonously towards negative values. Conversely, for negative
pions, the flow signal becomes at first positive, reaches a peak at
about $p_t$=0.01-0.02 GeV/c, and then decreases towards negative values
where the flow signals for both pion charges merge.  This is best seen
in Fig.~\ref{fpipm}.  The merging point appears to depend on centrality,
moving to lower $p_t$ for the more central events.  ii) An intermediate
region, approximately at $0.1<p_t<0.3$~GeV/c, where $v_1$ is negative
and only weakly dependent on $p_t$ for both pion charges.  And finally
iii) the high $p_t$ region where $v_1$ begins to rise and becomes
positive. There is in fact a very systematic rapidity dependence of the
zero-crossing point: It moves to lower $p_t$ values with increasing
rapidity occurring at $p_t$ = 0.5-0.6, 0.4-0.5, and 0.3-0.4 GeV/c for the
three rapidity bins at $y$=2.8-3.2, 3.2-3.6, and 3.6-4.0, respectively.

The centrality dependence of all flow signals is rather pronounced; in
the analyzed centrality region the magnitude of flow for all particles
decreases for more central collisions.  Directed flow of beam rapidity
protons shows a relatively smaller centrality dependence than that of
protons at lower rapidities (see also Fig.~\ref{fmpx} below).
 
\subsection*{Higher Harmonics in Azimuthal Distributions}

Higher order harmonics ($v_2$ and $v_3$) in the particle azimuthal
distributions have also been analyzed.  The accuracy in the evaluation
of the contribution of higher harmonics deteriorates with increasing
order due to the finite reaction plane resolution~(see the discussion
in~\cite{l877flow2}).  This results in larger relative errors.  Our
results for the proton elliptic flow (amplitude of the second harmonic
in the proton azimuthal distributions) as a function of transverse
momentum are presented in Fig.~\ref{fv2prot}.  A clear positive signal
is observed in the high $p_t$ region for rapidities of 2.6 and
larger. There is an indication that the signal moves to higher $p_t$
values with decreasing rapidity.  This combined with the smaller
acceptance in $p_t$ at rapidities below 2.6 may be the reason that no
significant signal is observed there.  In our previous measurements of
transverse energy and charged particle flow ~\cite{l877flow2} a clear
signal of elliptic flow was observed at all values of
pseudorapidity. The observed positive values of $v_2$ correspond to an
elliptically shaped distribution with the major axis lying in the
reaction plane. This orientation of the elliptic component is
perpendicular to what was measured at lower beam energies
\cite{squeeze}.

Within our spectrometer acceptance, pions do not exhibit
elliptic flow at the level of 2--3\%. 

The (absolute) accuracy in measuring  $v_3$ is about 0.1.
For all particles, rapidities, and centralities of the collision
the observed signals are consistent with zero within this accuracy.

\subsection*{Mean Directed Transverse Momentum}

Weighted with $p_t$ and its probability distribution the coefficient
$v_1$ yields $\mpx$, the mean value of the transverse momentum projected
onto the reaction plane:
\be
\mpx = \frac{1}{N} \int  v_{1}(p_t) p_{t}  \frac{d N}{d p_{t}} d p_{t}.
\label{empx}
\ee
Like any other integral quantity, $\mpx$ contains less information than
$v_1$. We nevertheless calculate this quantity in order to compare our
results with results from other experiments and model predictions.  This
is done by using spectra~\cite{l877pid} measured with the same
apparatus.  Our results of the value of $\mpx$ for protons at different
centralities are shown in Fig.~\ref{fmpx}.  Due to the experimental
acceptance in $p_{t}$, a model-independent evaluation of this quantity
is possible only at rapidities $y>2.8$.  Where it becomes necessary, we
extrapolate $dN/dp_t$ to high $p_t$ using a thermal parameterization (as
used e.g. in Equation (\ref{efun3})).  The filled points in Fig.~\ref{fmpx}
correspond to $\mpx$ calculated in accordance with Equation
(\ref{empx}), using the parameterization of $v_1(p_t)$ shown in
Fig.~\ref{fv1prot} as the solid (upper) curves to extrapolate $v_1$ into
the $p_t$ range not measured (for an analytic expression of this
parameterization see Equation~(\ref{ev1mts}) below).  The small
difference between the parameterization and the data in the low $p_t$
region does not contribute visibly to the final result for $\mpx$.  The
contribution of the non-measured high $p_t$ part of the spectra to
$\mpx$ is relatively small at rapidities $y\ge2.8$ (less than 10\%),
where the acceptance in $p_t$ is large.  At rapidity $y=2.5$ this
contribution accounts for about 40\% of the value of $\mpx$.  The
error-bars shown do not include the uncertainty in the effective slope
parameters used for the extrapolation of the $p_t$ spectra or any
systematic uncertainty associated with the extrapolation of $v_1$.

In order to assess the systematic error we evaluate the same quantity
($\mpx$) by using a different  
parameterization of the invariant triple differential distribution
with $\mpx$ as a parameter. We use the following functional form  to
parameterize  $d^{3}N / d^3p$ of protons in a given rapidity bin:
\bea
d^{3}N / d^3p & \propto & 
m_t' \exp\left(-\frac{m_t'-m}{T}\right), 
\label{efunc1}
\\
& & 
m_t'=\sqrt{(p_{x} - \langle p_{x}\rangle)^{2}+p_{y}^{2}+m^{2}},
\eea
{\it i.e.} a thermal distribution with respect to an origin displaced
along the $p_x$-axis. With the effective slope parameters taken
from~\cite{l877pid} we use 
function (\ref{efunc1}) to fit the experimental values of $v_1(p_t)$
(see Fig.~\ref{fv1prot}, dashed (lower) curves).  The extracted values
of $\mpx$ are shown in Fig.~\ref{fmpx} as open symbols.  The difference
between the results obtained with the two different parameterizations
(filled and open symbols in Fig.~\ref{fmpx}) gives an idea of the
systematic uncertainty of the results, not including the systematic
uncertainty in $v_1$ itself in the range where it is measured (see
above).  The latter could be important for the lowest rapidity region of
$2.2<y<2.4$, where the flow signal is very small; there we estimate an
associated systematic uncertainty of $\mpx$ of about 20~MeV.

The evaluation of $\mpx$ values for pions was done by convoluting the
experimental values of $v_1(p_t)$ with the spectra without any
extrapolation to the high $p_t$ region.  This is possible due to the
relatively large $p_t$ acceptance for pions in the rapidity interval
studied.  The extracted values of $\mpx$ for pions are about an order of
magnitude smaller than those for protons.  The results for the
centrality region $\sigma_{top}/\sigma_{geo}\approx$ 9--13\% (centrality
2) are presented in Fig.~\ref{fmpx2} (squares and triangles) together
with the corresponding proton results and will be discussed below.

\section{Directed flow discussion}
\label{sdisc}

\subsection*{Moving thermalized source {\em vs.} absorption}

One of the simplest pictures of directed flow would be the motion in the
transverse plane of a thermalized source, localized in
rapidity. Assuming for the simplest case of no radial expansion, the
invariant momentum spectrum for particle emission from a thermalized
source is described by:
\be
\frac{1}{m_t}\frac{dN}{dm_t dy d\phi} \propto E^* e^{-E^*/T},
\label{edn}
\ee
where $T$ is the temperature, and $E^*$ is the particle's energy in the
rest frame of the source.  In the case of a thermal source moving in
$x$-direction with velocity $\beta_{x}$ (which we call the directed flow
velocity), the value of $E^*$ can be obtained by a Lorentz
transformation:
\be
     E^*=\gamma \tilde{E} - \beta_x \gamma_x p_t \cos(\phi), 
\label{eet}
\ee
where $\gamma_x= 1/\sqrt{1-\beta_x^2}$, and the energy $\tilde{E}$ is
evaluated in the system moving longitudinally with the same velocity as
the source, $\tilde{E}=m_t \cosh(y-y^*)$.  Here $y^*$ is the source
rapidity.  Using Equations (\ref{edn}) and (\ref{eet}) one can evaluate
the first Fourier coefficient $v_1$ by direct integration to yield
\be
 v_1 = \frac{I_1(\xi)-(\xi I_0(\xi)-I_1(\xi))T/\tilde{E}}
       {I_0(\xi)-\xi I_1(\xi)T/\tilde{E}},
\label{ev1}
\ee
where $\xi=p_t \beta_x \gamma_x /T$, and $I_0(\xi)$, $I_1(\xi)$ are 
the modified Bessel functions.
For protons, $m_t\gg T$ and $v_1$ does not depend on the, generally unknown,
value of $y^*$.
Since we expect that $\beta_x \ll 1$ it is useful to note that in that
case, for relatively
small values of $p_t$, $v_1$ depends almost linearly on $p_t$:
\be
v_1 \approx \frac{I_1(\xi)}{I_0(\xi)} \,\, \,\, 
\left( \approx \frac{p_t \beta_x}{2T} \right).
\label{ev1mts}
\ee
The solid lines shown in Fig.~\ref{fv1prot} correspond to fits to the
data using Equation (\ref{ev1mts}) and inverse slope parameters $T_{B}$
from~\cite{l877pid} as $T$. Overall, the fit is rather good. Taking
into account that, in general, one cannot interpret the inverse slope
parameters $T_{B}$ as a source temperature we would like to note that
Equations~(\ref{ev1},\ref{ev1mts}) are still valid for the case that the
invariant spectra at a fixed rapidity have a thermal shape with some
effective temperature constant.

In Fig.~\ref{fbetax}, the extracted values of $\beta_{x}$ of the proton
source are shown for different collision centralities as a function of
proton rapidity. One can see that the transverse source velocity grows
with increasing rapidity, possibly peaking around rapidity y=3.1. The
maximum values are about 10 \% of the speed of light. This transverse
velocity is found to decrease with increasing centrality of the
collision. These results have to be put into perspective with the
transverse expansion velocities fitted to spectra of various particle
species in the same reaction and close to mid-rapidity: in such an
analysis transverse expansion velocities of about 50 \% of the speed of
light are required to describe the data\cite{lstachel}. This effect is
largest in central collisions. The transverse expansion velocity
decreases away from mid-rapidity.

Looking more closely at the fits in Fig.~\ref{fv1prot} one notices a
small deviation of the fit using Equation (\ref{ev1mts}) from the data
at small transverse momenta ($p_t<0.1$~GeV/c) and at rapidities $y\le
2.8$.  There, protons exhibit a weaker or even opposite flow than
expected from the simple model discussed above.  This could be an
indication of transverse expansion as pointed out in
reference~\cite{lsvrd}. There, it is shown that, depending on the
relative magnitude of the sideward flow velocity $\beta_x$ and the
transverse expansion velocity $\beta_t$, a reduction or even sign change
of $v_1$ at small $p_t$ is possible.

For pions we observe a dependence of $v_1$ on $p_t$ very different from
that for protons.  It implies that the physics of pion flow is different
from that of a moving source alone (just as one would expect).  A
possible explanation of the pion flow signal could be found in a
superposition of different effects: Absorption in nuclear matter,
Coulomb interaction with comoving protons, and sideward motion of the
source.

For pions produced in the center of the collision volume and moving
in forward direction with velocities close to speed of light, one
would expect more nuclear matter on the side characterized by the
direction of nucleon flow (positive $x$ in our notation) and,
consequently, more absorption.  The effect of pion absorption on
nucleons with energy-, or for a fixed rapidity, $p_t$-independent cross
section would lead to a 
negative and constant value of $v_1(p_t)$.  Such absorption should be
comparable for positive and negative pions.  On the other hand, it is
known that in the relevant momentum range the pion-nucleon elastic and
total cross sections are strongly peaked at the $\Delta$
resonance~\cite{lprd}. In a frame where the nucleon is at rest the cross
section peaks at a pion momentum of 0.3 GeV/c and falls off rapidly for
larger momenta. If the absorption of pions would occur on nucleons of the same
rapidity, this would mean that the negative values of
$v_1$ should increase towards zero for $p_t \geq$ 0.3 GeV/c. Most of the
pions we observe are in a rapidity range where there are very few
nucleons (forward of beam rapidity). Indeed, analysis of the proton
rapidity distribution indicates~\cite{lstachel} that nucleon sources
(fireballs) are distributed evenly over plus and minus one unit of
rapidity in the c.m. frame, {\it i.e.} that the most forward nucleon
source is at rapidity 2.5 in the laboratory. For pions at rapidities
3.0, 3.4, and 3.8 this would imply that absorption effects should
decrease for transverse momenta larger than 0.30, 0.27, and 0.20
GeV/c. The data indeed show this trend with rapidity at about the $p_t$
values expected (see Figs.~\ref{fv1pp},\ref{fv1pm}, and \ref{fpipm}).

At the same time Coulomb interaction with positively charged nuclear
matter would be different for pions of different charge, and could
qualitatively explain the difference in flow signals at very low $p_t$
shown in Fig.~\ref{fpipm}. Negative pions are attracted to the protons
leading to positive values of $v_1$ (as argued in the case of absorption
on nucleons above) while positive pions are repelled leading to more
negative values of $v_1$.

The rise of $v_1$ for pions towards large $p_t$ could reflect the sideward
motion of the source, the same way it does for protons. 

\subsection*{Comparison with the results on $N_c$ and $E_T$ flow}

Fig.~\ref{fmpx2} shows, for the $E_T$-range 200-230 GeV, the mean
directed transverse momenta for protons, positive, and negative
pions. Due to the acceptance of the spectrometer the proton and pion
results are forward of rapidity 2.2 and 2.8, respectively. Results were
obtained using Equation~(\ref{ev1mts}) to complete the integral
(\ref{empx}) beyond the range where $v_1$ is measured. The
error-bars shown include an error in the effective temperature of about
7 \% in addition to the statistical errors. The symmetry of the
collision system allows reflection of these results about mid-rapidity,
shown in Fig.~\ref{fmpx2} by the open symbols, providing values for
protons backward of rapidity 0.9 and for pions backward of rapidity
0.3. To fill the gap at intermediate rapidities we use our measurement
of the flow signal in transverse energy $E_T$ and charged particle
multiplicity $N_c$ as published in ~\cite{l877flow2}. In this paper
transverse energy and charged particle flow were decomposed into nucleon
and pion flow under a few simple assumptions.  Both were studied as a
function of pseudorapidity.  Using these results and a simple
parameterization of the proton and pion spectra in $y$ and $p_t$ from
the AGS experiment E866~\cite{lakiba} and our own
measurement~\cite{l877pid} which together provide complete rapidity
coverage, we evaluate $\mpx$ as a function of rapidity.  More
specifically, we parameterize the spectra by expressions of the form:
\bea
\frac{d^2N}{m_t dm_t dy}&\propto& \exp(-\frac{(y-y_0)^2}{2\cdot\sigma_y^2})
\cdot m_t \, \exp(-\frac{m_t-m}{T_b(y)}),
\eea
and
\bea
& &T_b(y)=T_0 \exp(-\frac{(y-y_0)^2}{2\cdot\sigma_T^2})
\eea
with values of $\sigma_y$ = 0.89 (0.88), $T_0$ = 0.26 (0.15) GeV, and
$\sigma_T$ = 1.06 (1.70) for protons (charged pions).  Using these
parameterizations we calculate the mean rapidity $y(\eta)$ (weighted
with $E_T$) and the width of the rapidity interval associated with a bin
in pseudorapidity. Then, under the assumption that $v_1^{E_T}$ and
$v_1^{p_t}$ are similar, one obtains $\mpx$ as a function of rapidity:
\be
\langle p_x(y) \rangle \approx \langle p_{t}(y) \rangle \cdot v_1^{E_T}(y). 
\ee
The resulting mean transverse momenta are presented as stars in
Fig.~\ref{fmpx2}, together with our spectrometer results, for protons
and pions; the errors shown are a propagation of the statistical errors
in the measured flow of $E_T$ and $N_c$.  The horizontal error-bars
shown with the stars represent the widths of the rapidity region
actually contributing to the pseudorapidity bins for which $v_1$ was
measured.

Overall, the agreement between the results of the current analysis of
spectrometer data at forward (and backward) rapidities and the results
derived from the measurements of $E_T$ and $N_c$ flow is good. For the
latter method we had estimated the systematic error for proton flow to
be about 10\% ~\cite{l877flow2}; this can also be judged from the
symmetry of the data shown in Fig.~\ref{fmpx2} with respect to
midrapidity (note that all stars correspond to measured points). The
dotted line has been added to aid inspection of the data for the
necessary antisymmetry. The first and the next to last proton point have
equal distance from mid-rapidity and should be identical. They differ by
10 MeV/c (out of 80 MeV/c). For the spectrometer results the systematic
errors are largest at the lowest rapidities shown (see Fig.~\ref{fmpx}
and discussion) where they are estimated to be 20 MeV/c.

Within these errors the points obtained from the two methods merge
smoothly and taken together they provide a measurement of $\mpx$ for
protons and charged pions as a function of rapidity over practically the
entire relevant rapidity range.  One can see a proton flow signal rising
away from midrapidity to maximal values of about 130 MeV/c close to beam
and target rapidities. The pion signal is in comparison very small,
about 1/20 or the proton signal, but nevertheless significant. It is
directed opposite to the nucleon flow signal.

\subsection*{Comparison with RQMD predictions}


In Fig.~\ref{fmpx2} we compare our results for $\mpx$ with the
predictions of the RQMD event generator~\cite{lrqmd,lrqmd23}, versions
1.08 and 2.3.  Version 1.08, run in cascade mode, has been used for
comparison to our earlier results for Au + Au collisions on
pseudorapidity distributions~\cite{l877dndeta}, proton and pion
rapidity~\cite{l877pid} and transverse mass~\cite{lstachel}
distributions as well as results for Si+A and p+A collisions. For the Au
+ Au system it was found that overall many features of the data are
predicted correctly with one striking deviation: generally proton
spectra are significantly too steep in the model; at midrapidity the
inverse slope constant is only 2/3 of the experimental
value. Predictions from this version of the model have also been
compared to our experimental results for flow in $E_T$ and
$N_c$~\cite{l877flow1,l877flow2} and generally the model exhibited too
little flow as compared to the data. Comparing to the proton data in
Fig.~\ref{fmpx2}, we find that the model also underpredicts the proton
flow by nearly a factor of two. For pions, the sign change of the flow
relative to protons is properly predicted but in magnitude the model
overpredicts this opposite pion flow by a factor of 2 to 3.

Meanwhile, a new version of RQMD (version 2.3) has been
developed~\cite{lrqmd23}. It has been known for some time~\cite{lmatie}
that taking into account mean field effects by simulating in the model a
Skyrme-type nucleon-nucleon potential increases the slopes of proton
spectra and at the same time increases the proton flow signal while
reducing the opposite pion flow signal in magnitude. We have used the
most recent version of the model in the so-called ``mean field'' mode to
compare to the present data as well. Indeed, in this version of the code
the proton flow signal is larger by about 50\% for the proton $\mpx$ at
the peak and the data are described rather well (see
Fig.~\ref{fmpx2}). Simultaneously, the opposite pion flow is reduced,
again to a level consistent with the experimental data. It should be
noted that at the same time the slope difference in the proton spectra
is reduced (at midrapidity the slope constant of the model is now 3/4 of
what is observed in the data) but a deviation persists at all
rapidities.

Fig.~\ref{fv1rqmd} shows that the agreement of the RQMD predictions
(version 2.3, mean field) with the experimental results for $\mpx$ of
protons as displayed in Fig.~\ref{fmpx2} may indeed be accidental. The
functional dependence of the proton flow signal, expressed now in terms
of the first Fourier coefficient $v_1$ as a function of $p_t$, is very
different from what is seen in the data. In the model, the flow signal
rises at first with increasing $p_t$ and then becomes rather flat. In
fact, in the rapidity bin around y = 2.9 the model overpredicts $v_1$
but at the same time underpredicts the overall slope of the proton
spectrum by 25 \% giving overall agreement in $\mpx$. An interesting
question arises whether the rapidly rising and saturating $p_t$
dependence as exhibited by the RQMD predictions (which is in fact a
general trend observed for all rapidity bins) is due to the
rescattering-type mechanism which produces flow in a cascade code in
contrast to a hydrodynamic flow mechanism (reflected {\it e.g.} by the
$p_t$ dependence of Equations~(\ref{ev1},\ref{ev1mts})).

The elliptic flow signal seen in the anisotropy of $E_T$ and $N_c$ was
found to be in good agreement with the RQMD (version 1.08) prediction in
our previous publication \cite{l877flow2}. As discussed above, the sign
of this elliptic flow signal changes from negative (preferential
emission perpendicular to the reaction plane) at lower beam energies
(1-2 GeV/nucleon range at the Bevalac and SIS) to a positive signal
(preferential in-plane emission) at the present energy. It was recently
noted by Sorge~\cite{lsorge1} that the $v_2$ anisotropies are sensitive
to the pressure at the maximum compression. Positive values of $v_2$
were in fact predicted in a hydrodynamic model by Ollitrault for very
high energies~\cite{lolli}. The observed change in sign between
Bevalac/SIS and AGS energies therefore implies that at AGS energies
already the final in-plane flow overwhelms the initial shadowing
effect. It would be interesting to deduce, from this information, the
pressure achieved at maximum baryon density in Au + Au collisions at the
AGS.

\section{Summary and Conclusion}

The measured flow signals of identified particles show that, in
semi--central collisions, protons (nucleons) exhibit strong directed
flow. Pions of low $p_t$ exhibit flow in the opposite direction; at
higher $p_t$ pions start to flow in the same direction as protons. The
results on $\mpx$ values for proton and pion are complementary in
acceptance and match well with nucleon and pion flow values derived from
the same experiment by measurement of $E_T$ and $N_c$
flow~\cite{l877flow2}.  Taken together, these results represent a
directed flow measurement of protons and pions over nearly the entire
rapidity region in Au+Au collisions at 11$A$ GeV/c. The flow is found to
be maximal around beam and target rapidities with values of $\mpx$ for
protons of about 130 MeV/c and 5\% of that for charged pions.

The observed positive values of the second harmonic amplitudes of the
proton azimuthal distributions correspond to preferential particle
emission in the reaction plane.  This observation agrees with our
previous measurements of $E_T$ and $N_c$ elliptic flow~\cite{l877flow2}.

The nearly linear dependence of the proton flow signal $v_1$ on $p_{t}$
can be interpreted in the framework of a transversely moving thermal
source.  The corresponding source velocity $\beta_x$ appears to reach
values of 0.1 in the beam rapidity region and for semi--central
collisions.  The more complicated $p_t$ dependence of the pion flow
signals indicates that the effect there is probably a superposition of
several effects such as absorption, Coulomb interaction and overall
motion of the source (as for protons). 

\section*{ACKNOWLEDGMENTS}

We thank the AGS staff, W. McGahern and Dr. H. Brown for excellent
support and acknowledge the help of R. Hutter in all
technical matters. Financial support from the US DoE, the NSF, the
Canadian NSERC, and CNPq Brazil is gratefully acknowledged. One of us
(JPW) thanks the A. v. Humboldt Foundation for support,
while another (WCC) acknowledges the financial support from 
the Gottlieb-Daimler and Karl-Benz-Stiftung during his 
stay in Heidelberg, Germany, where part of this work was done.
\section*{Appendix}

\subsection*{Flattening of the Reaction Plane Distribution}

We apply the following procedure to correct for the  non-flatness 
of the reaction plane distribution. 
Any flattening of the distribution means a correction to 
the reaction plane angle.
We explicitly introduce this correction defining a new angle as:
\be
\Psi_1'=\Psi_1+\Delta \Psi_1.
\ee
where $\Delta \Psi_1$ is written in the form:
\be
\Delta \Psi_1= \sum_n ( A_{n} \cos(n \Psi_1) + B_{n} \sin(n\Psi_1))
\ee
Requiring the vanishing of the $n$-th Fourier moment of the new
distribution, the coefficients $A_{n}$ and $B_{n}$ can be evaluated by the
original distribution.
\bea
        B_{n} & = & \frac{2}{n} \la \cos(n\Psi_1) \ra,
\\
        A_{n} & = & -\frac{2}{n} \la \sin(n\Psi_1) \ra,
\eea
where the brackets refer to an average over events.
This gives:
\be
\Psi_1' = \Psi_1 + \sum_n \frac{2}{n} ( \;-\la\sin(n\Psi_1)\ra \cos(n\Psi_1) +
\la\cos(n\Psi_1)\ra \sin(n\Psi_1))
\ee

In practice, we flatten the reaction plane distribution up
to the fourth  harmonic (n=4).
Note that, due to the small values of $A_n$ and $B_n$ (typically of 
the order of a few percent), such a flattening of the distribution
does not have any effect on the reaction plane resolution.
It can also be shown that the same flattening procedure removes 
possible trigger biases (due to imperfect calibration,
dead channels or any other asymmetry) at least up to the second order.

\clearpage
\newpage
\section*{Figure captions}
\begin{enumerate}

\item \label{fe877}
  The E877 apparatus.

\item \label{fspectra}
  Proton $m_t$ distributions in the rapidity interval $2.8<y<2.9$
  for different centralities (indicated by an $E_T$ range in GeV)
  together with a fit using the  
  function given in Equation~(\ref{efun3}). The inverse slope values and
  their statistical errors are shown in GeV. 
  The open (filled) squares correspond to 
  $-\pi/4<\phi<\pi/4$ ($3\pi/4<\phi<5\pi/4$).

\item \label{fteff}
  The dependence of the inverse (Boltzmann) slope $T_B$ 
  of the proton distributions as a function of azimuthal angle.
  The open points are reflections of the filled points about $\phi=0$.

\item \label{fv1prot}
  The transverse momentum dependence of the first 
  moment ($v_1$) of the proton azimuthal distributions for different particle 
  rapidities and centralities of the collision.
  The solid and dashed curves are fits using functions given in
  Eqs.~(\ref{ev1mts}) and~(\ref{efunc1}), respectively. For a description see
  Sections~\ref{sresu} and \ref{sdisc}.

\item \label{fv1pp}
  Same as Fig.~\ref{fv1prot} for $\pi^+$.

\item \label{fv1pm}
  Same as Fig.~\ref{fv1prot} for $\pi^-$.
  The open and filled symbols correspond to data from the 1993 and 1994
  runs.

\item \label{fpipm}
  Fourier coefficients $v_1$ for pions in the low $p_t$ region
  for the rapidity interval $3.2<y<3.6$ and two different centralities.
  Solid symbols represent data for positive pions. 
  Open symbols are for negative pions (circles and squares correspond to
  results from the 1993 and 1994 runs, respectively).
  
\item \label{fv2prot}
  The transverse momentum dependence of the second 
  moment ($v_2$) of the proton azimuthal distributions for different particle 
  rapidities and centralities of the collision.
 
\item \label{fmpx}
  The mean projection $\mpx$ of the proton transverse momentum onto
  the reaction plane as a function of rapidity for different
  centralities of the collision. Results for the second and fourth
  centrality bin are 
  shifted to the left by 0.025 for clarity of the picture.
  Solid and open symbols correspond to fits of $v_1(p_t)$
  with functions given in Eqs.~(\ref{ev1mts}) and~(\ref{efunc1}), respectively.

\item \label{fbetax}
  The transverse velocity $\beta_{x}$ of a thermal (proton) source for 
  different rapidities and centralities of the collision. For details
  see text.

\item \label{fmpx2}
  Mean projection $\mpx$ of the proton and pion transverse momentum onto
  the reaction plane as a function of rapidity for the centrality
  bin with $E_T$ = 200-230 GeV. Solid circles, squares, and triangles
  correspond to measurements in 
  the spectrometer. Open circles, squares, and triangles are reflections of
  the filled symbols about midrapidity. The results derived from
  measurements of $E_T$ and $N_c$  
  flow~\cite{l877flow2} for nucleons and pions are shown by stars~(see text).
  The dotted line is added to aid
  inspection of the solid stars for antisymmetry about mid-rapidity. 
  Results from calculations using 2 versions of RQMD (1.08 in cascade mode,
  2.3 in mean field mode) are depicted as histograms.

\item \label{fv1rqmd}
  Comparison of measured values of $v_1$ for protons in two rapidity
  bins and the centrality
  bin with $E_T$ = 200-230 GeV with those predicted by 2 versions of the
  event generator RQMD. 

\end{enumerate}


\clearpage
\newpage

\begin{figure}
\centerline{\psfig{figure=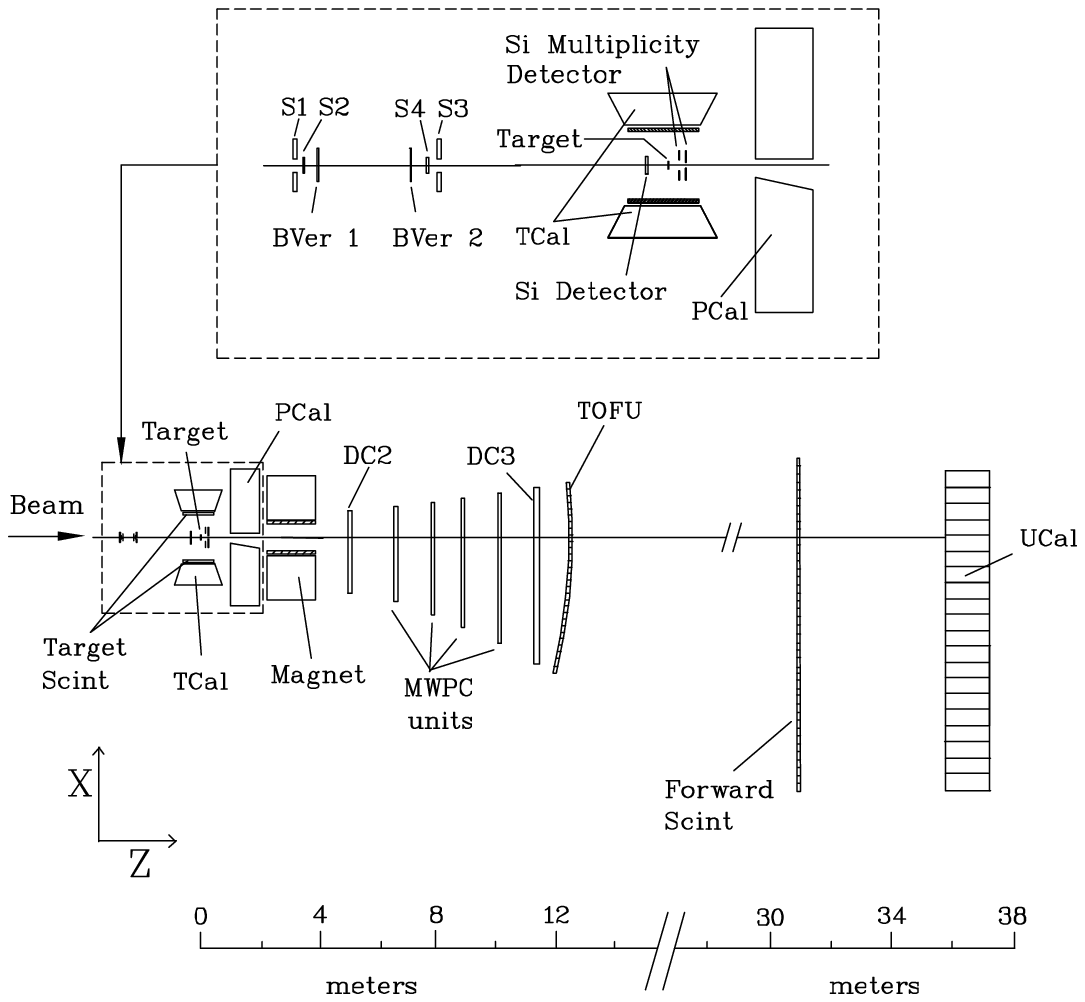,height=15.0cm}}
  \caption[]{
  The E877 apparatus.
    }
\end{figure}

\begin{figure}
\centerline{\psfig{figure=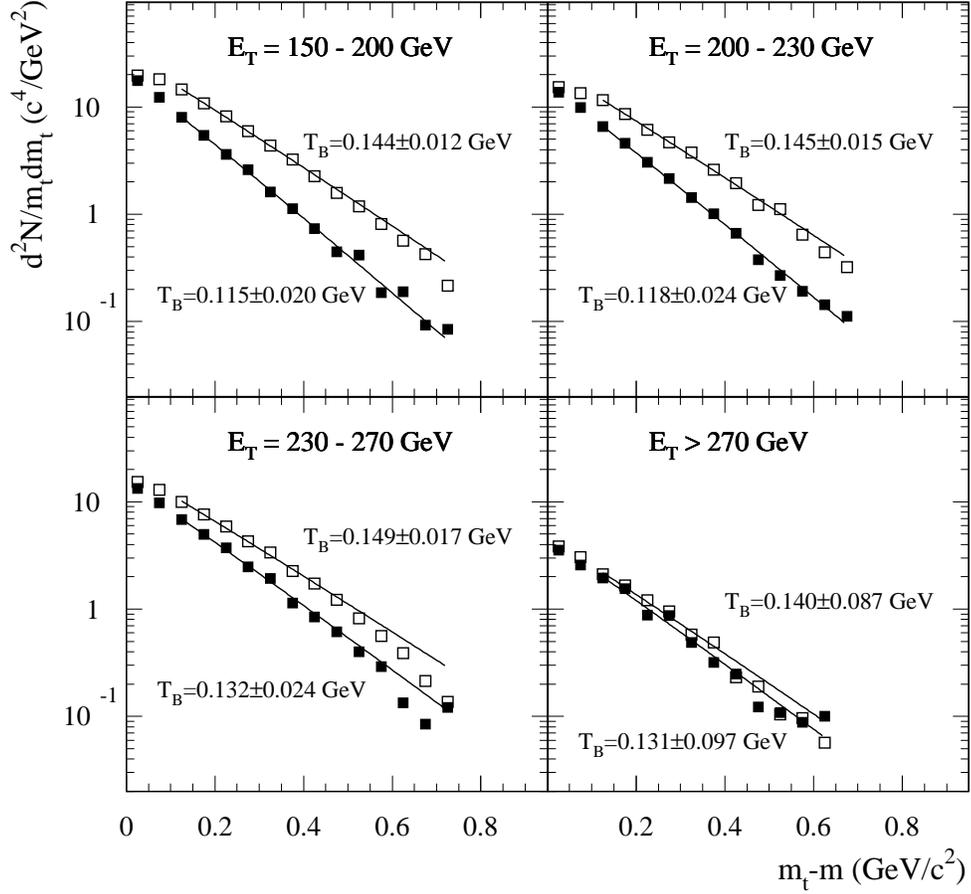,height=14.0cm}}
  \caption[]{
 Proton $m_t$ distributions in the rapidity interval $2.8<y<2.9$
  for different centralities (indicated by an $E_T$ range in GeV)
  together with a fit using the  
  function given in Equation~(\ref{efun3}). The inverse slope values and
  their statistical errors are shown in GeV. 
  The open (filled) squares correspond to 
  $-\pi/4<\phi<\pi/4$ ($3\pi/4<\phi<5\pi/4$).
    }
\end{figure}

\begin{figure}
\centerline{\psfig{figure=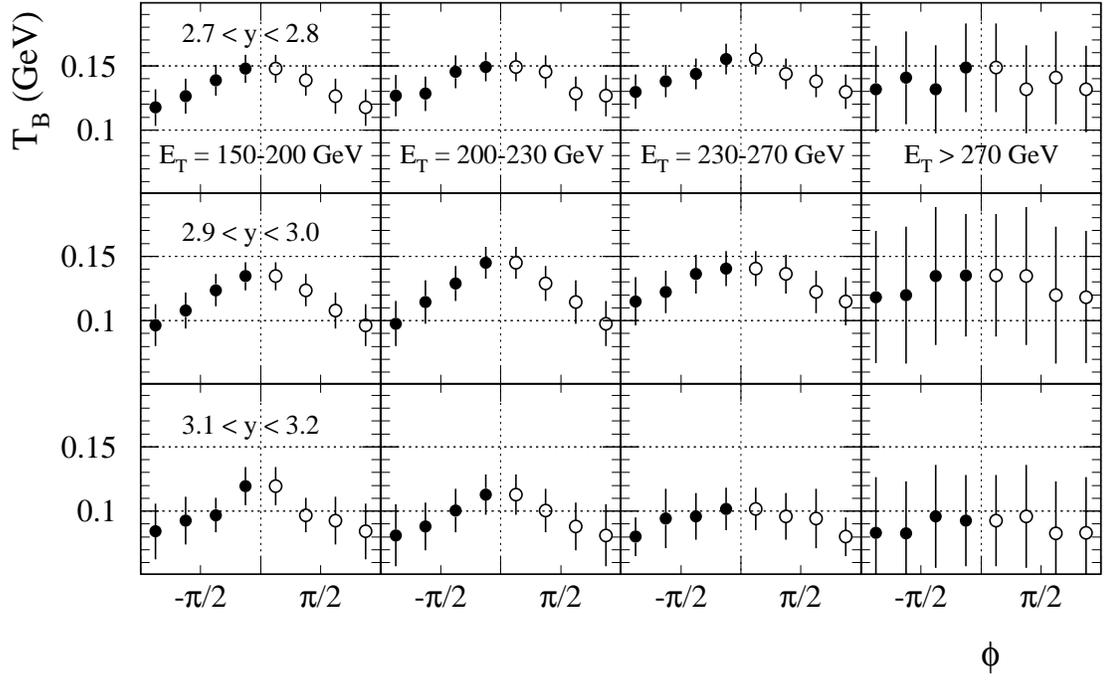,height=12.0cm}}
  \caption[]{
  The dependence of the inverse (Boltzmann) slope $T_B$ 
  of the proton distributions as a function of azimuthal angle.
  The open points are reflections of the filled points about $\phi=0$.
    }
\end{figure}

\begin{figure}
\centerline{\psfig{figure=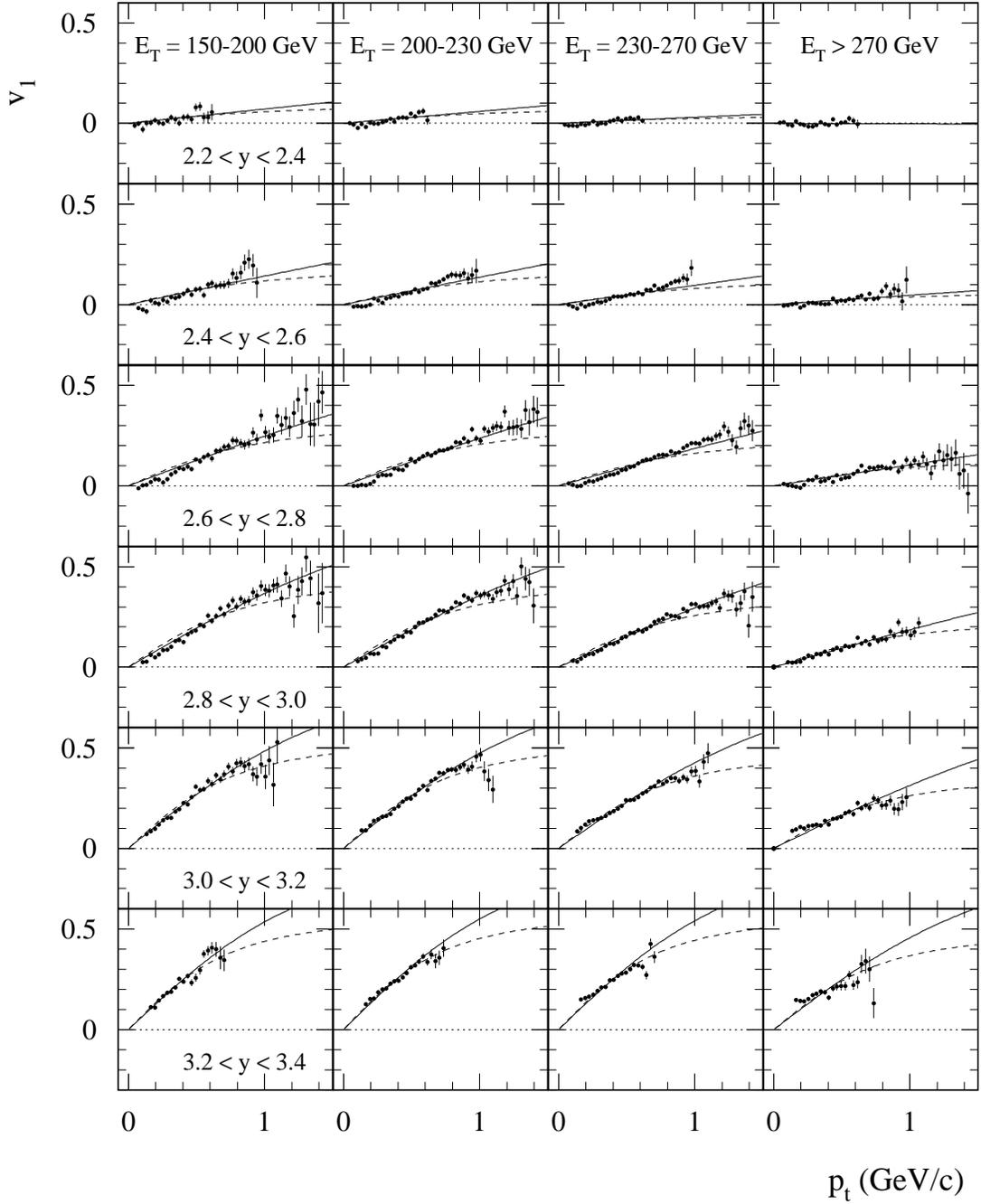,height=20.0cm}}
  \caption[]{
  The transverse momentum dependence of the first 
  moment ($v_1$) of the proton azimuthal distributions for different particle 
  rapidities and centralities of the collision.
  The solid and dashed curves are fits using functions given in
  Eqs.~(\ref{ev1mts}) and~(\ref{efunc1}), respectively. For a description see
  Sections~\ref{sresu} and \ref{sdisc}.
    }
\end{figure}

\begin{figure}
\centerline{\psfig{figure=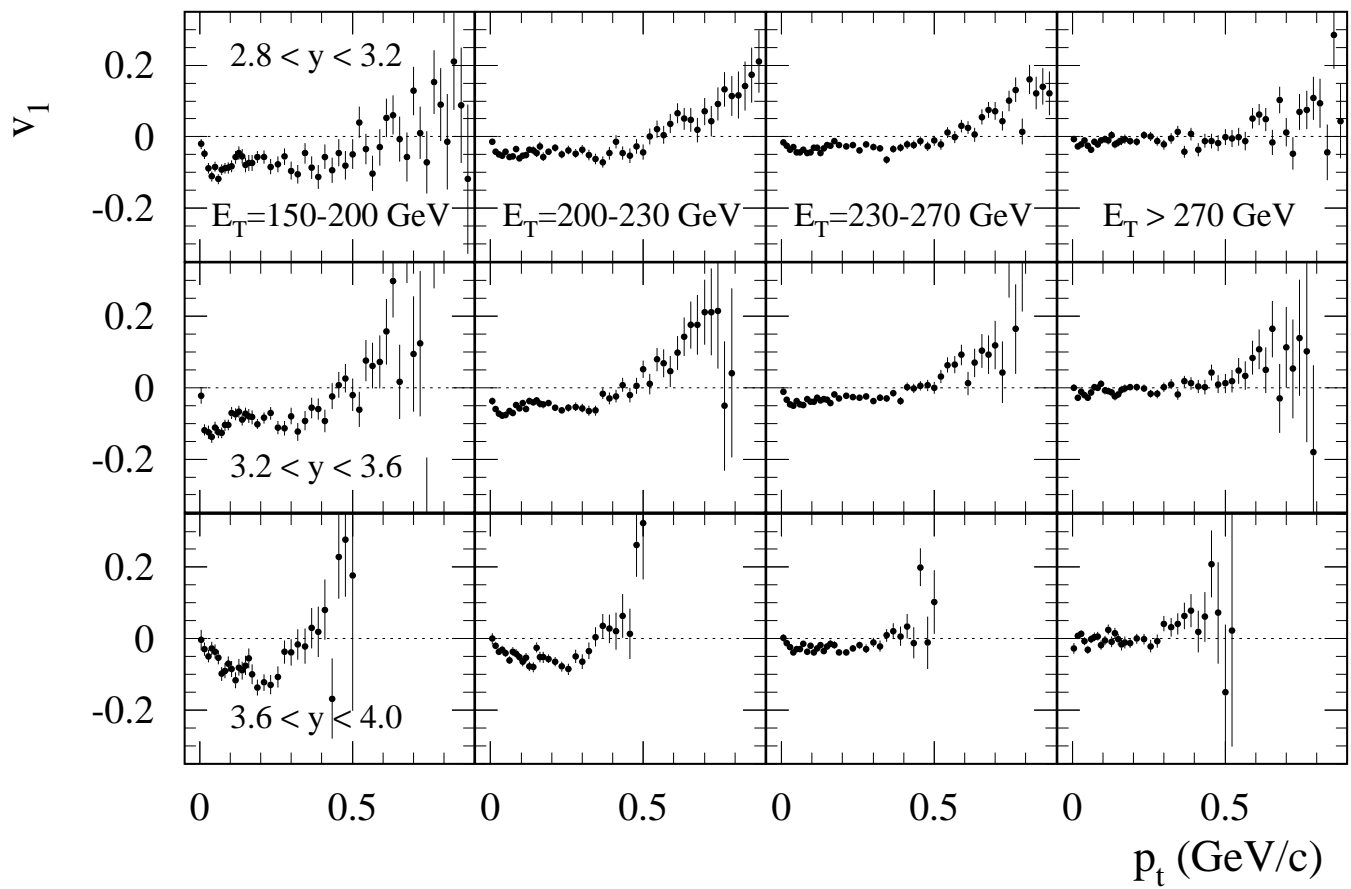,height=15.0cm}}
  \caption[]{
  Same as Fig.~\ref{fv1prot} for $\pi^+$.
    }
\end{figure}

\begin{figure}
\centerline{\psfig{figure=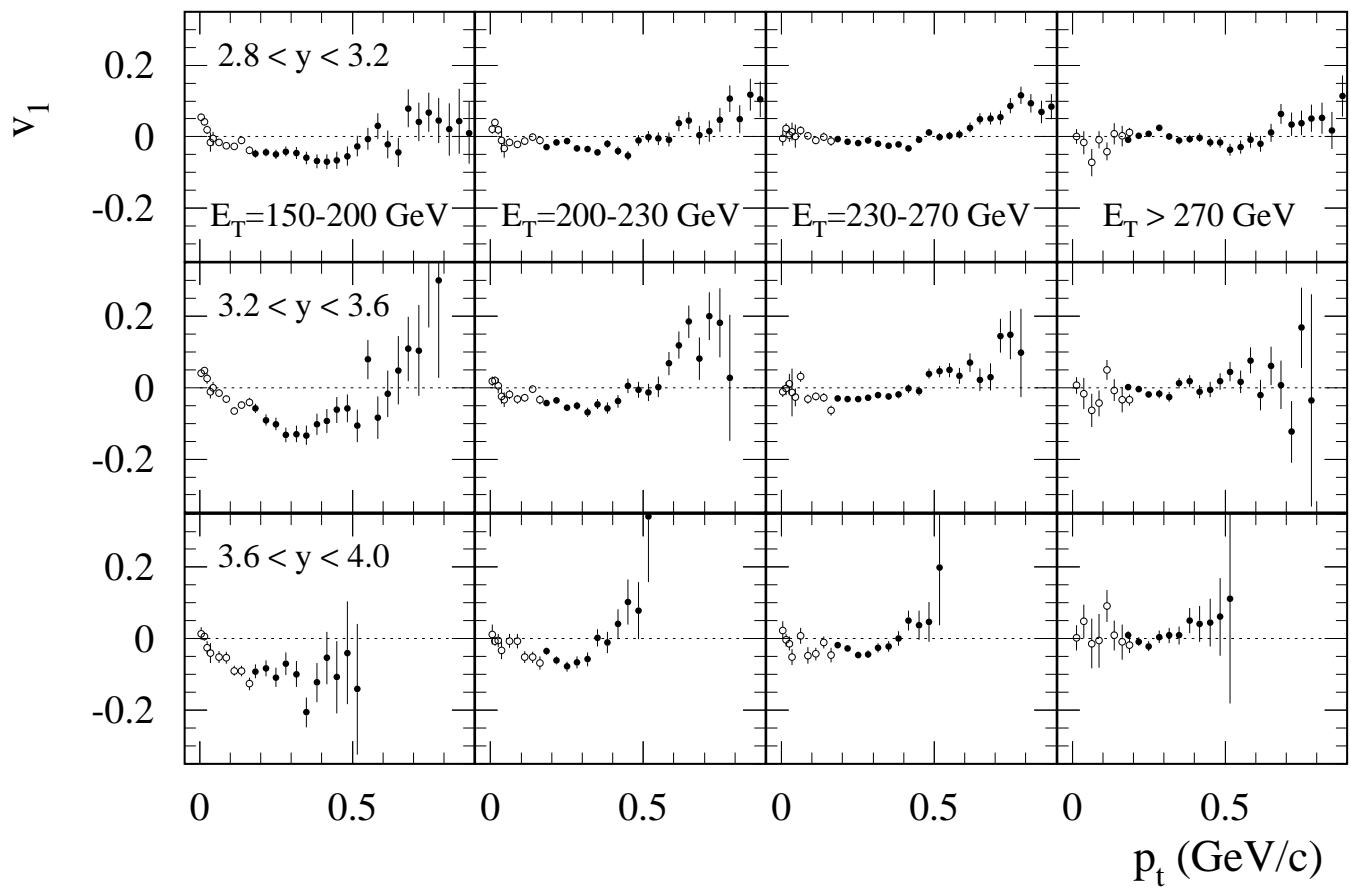,height=15.0cm}}
  \caption[]{
  Same as Fig.~\ref{fv1prot} for $\pi^-$.
  The open and filled symbols correspond to data from the 1993 and 1994
  runs.
    }
\end{figure}

\begin{figure}
\centerline{\psfig{figure=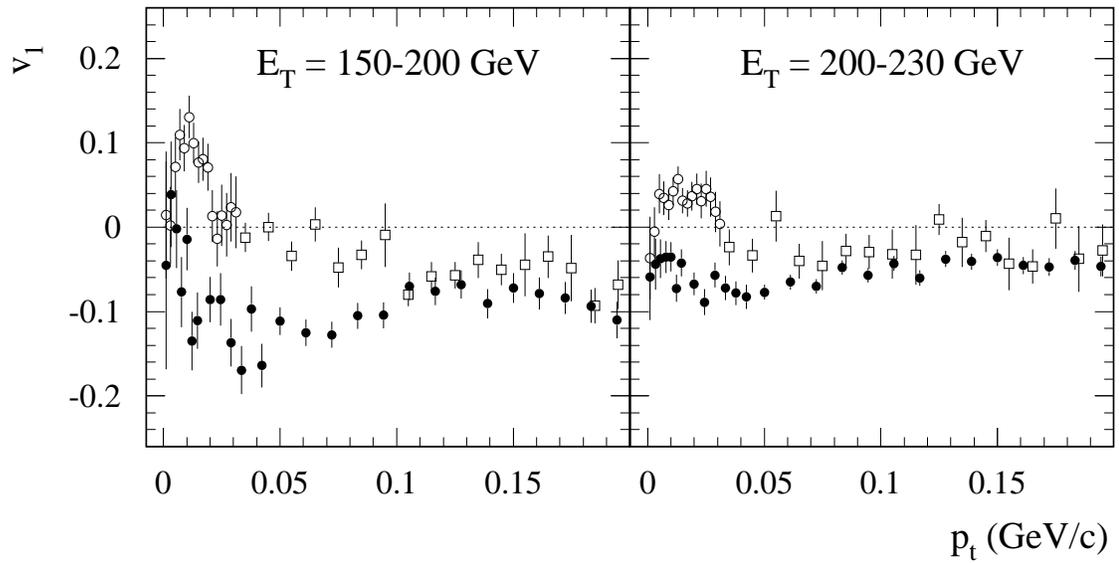,height=10.0cm}}
  \caption[]{
  Fourier coefficients $v_1$ for pions in the low $p_t$ region
  for the rapidity interval $3.2<y<3.6$ and two different centralities.
  Solid symbols represent data for positive pions. 
  Open symbols are for negative pions (circles and squares correspond to
  results from the 1993 and 1994 runs, respectively).
    }
\end{figure}

\begin{figure}
\centerline{\psfig{figure=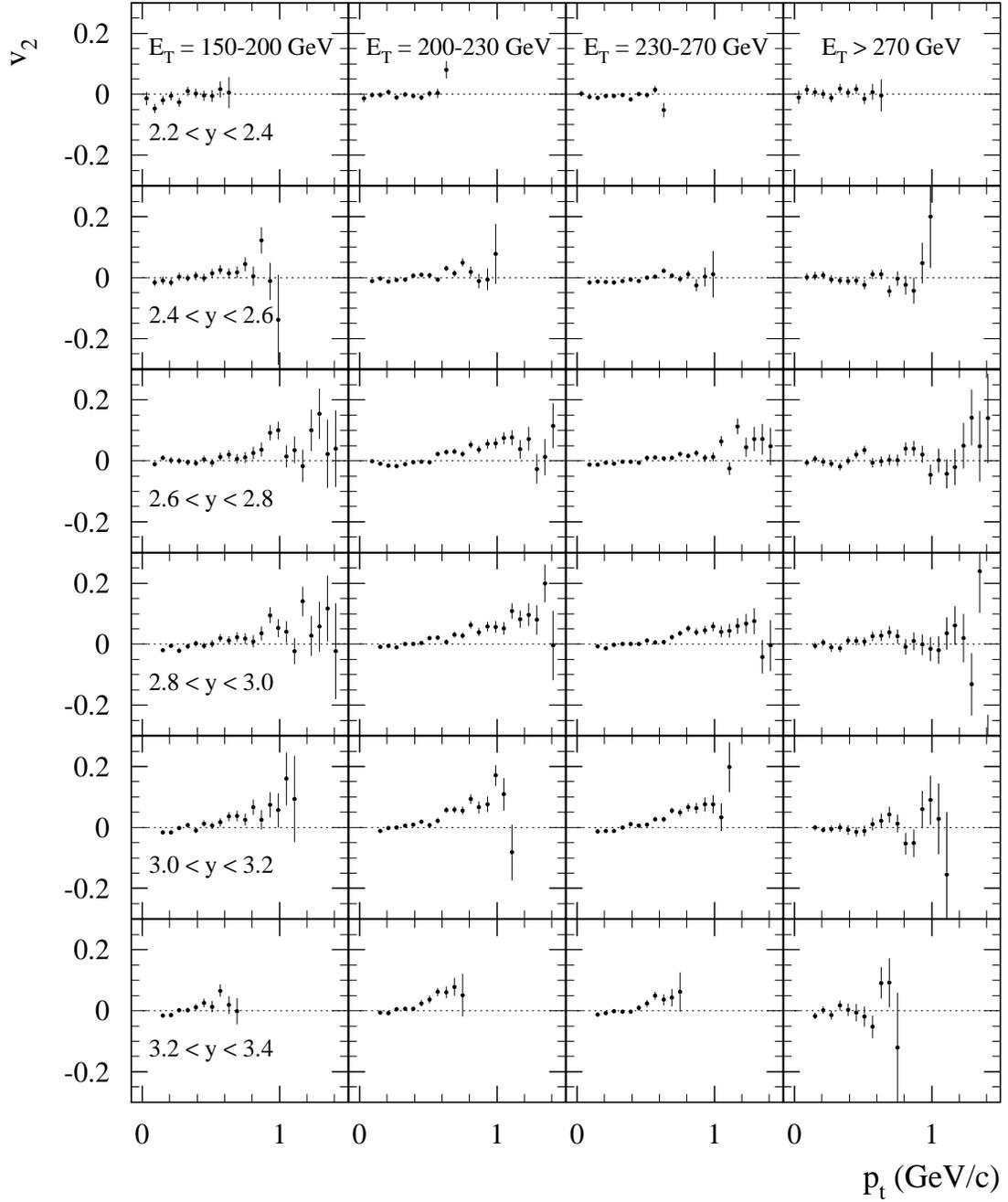,height=20.0cm}}
  \caption[]{
  The transverse momentum dependence of the second 
  moment ($v_2$) of the proton azimuthal distributions for different particle 
  rapidities and centralities of the collision.
    }
\end{figure}

\begin{figure}
\centerline{\psfig{figure=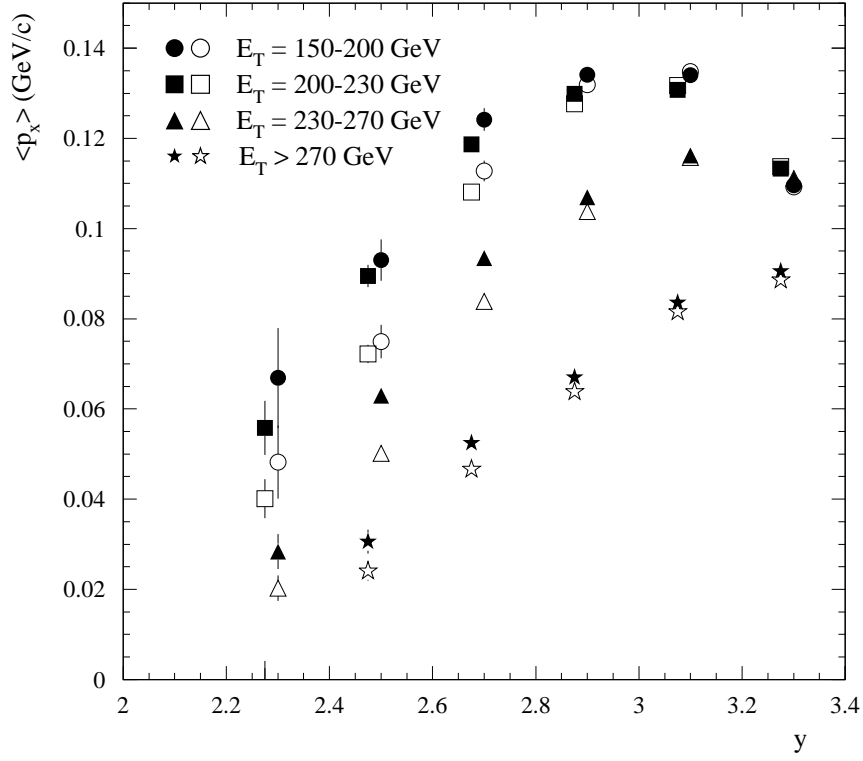,height=12.0cm}}
  \caption[]{
  The mean projection $\mpx$ of the proton transverse momentum onto
  the reaction plane as a function of rapidity for different
  centralities of the collision. Results for the second and fourth
  centrality bin are 
  shifted to the left by 0.025 for clarity of the picture.
  Solid and open symbols correspond to fits of $v_1(p_t)$
  with functions given in Eqs.~(\ref{ev1mts}) and~(\ref{efunc1}), respectively.
    }
\end{figure}

\begin{figure}
\centerline{\psfig{figure=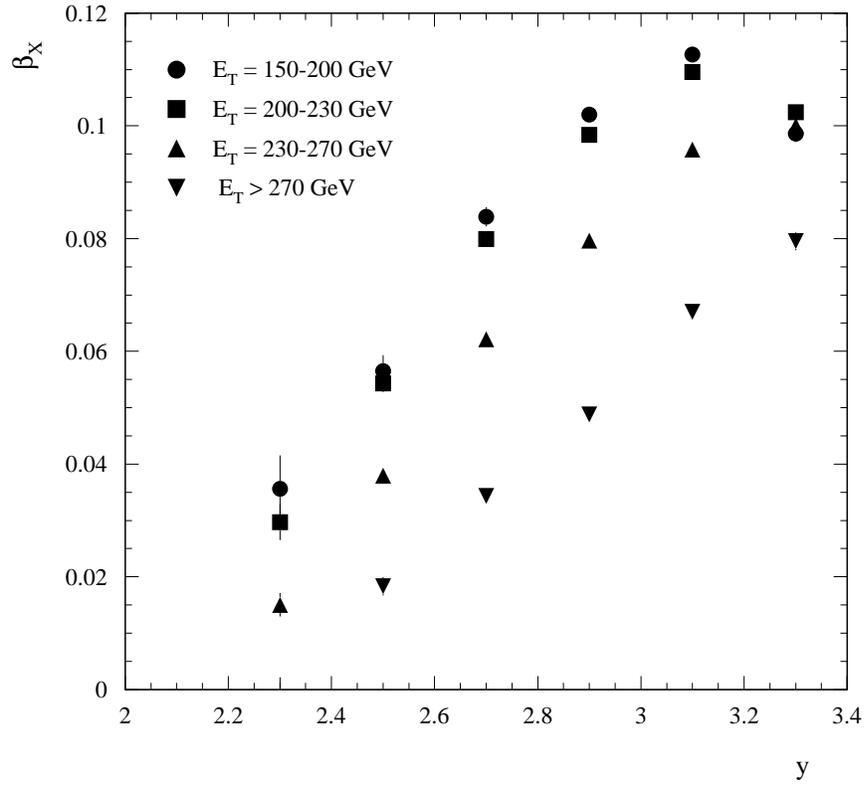,height=12.0cm}}
  \caption[]{
  The transverse velocity $\beta_{x}$ of a thermal (proton) source for 
  different rapidities and centralities of the collision. For details
  see text.
    }
\end{figure}

\begin{figure}
\centerline{\psfig{figure=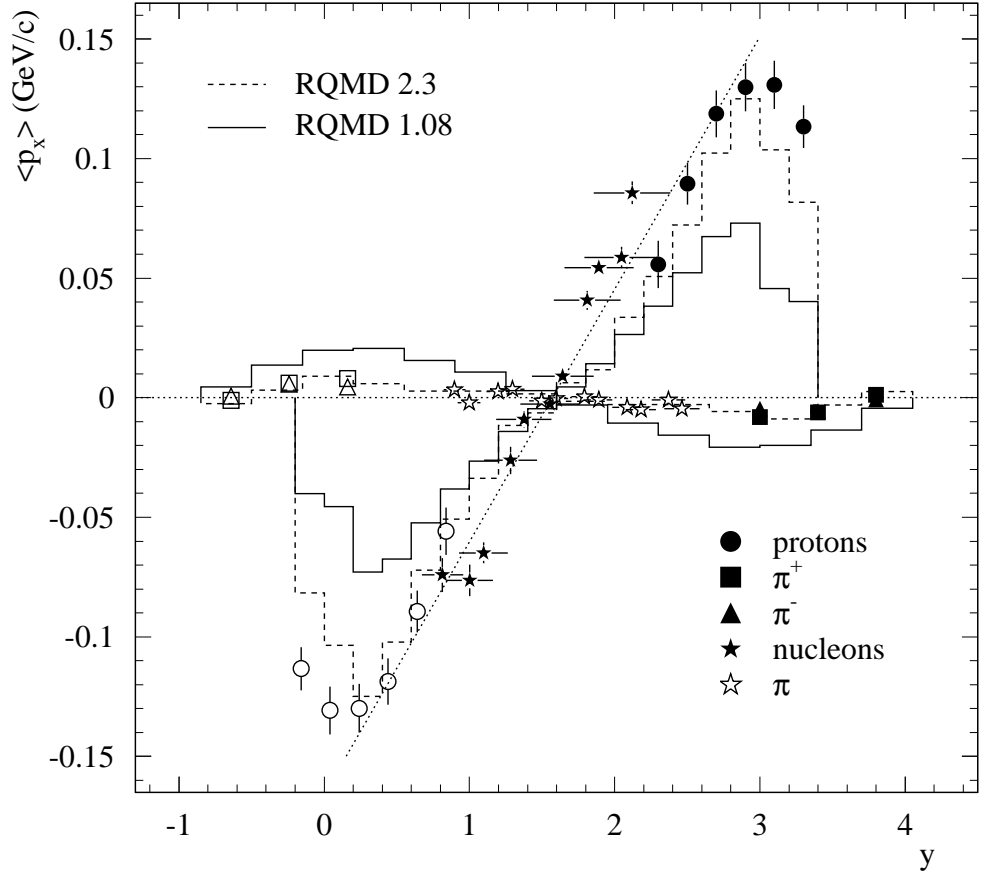,height=14.0cm}}
  \caption[]{
  Mean projection $\mpx$ of the proton and pion transverse momentum onto
  the reaction plane as a function of rapidity for the centrality
  bin with $E_T$ = 200-230 GeV. Solid circles, squares, and triangles
  correspond to measurements in 
  the spectrometer. Open circles, squares, and triangles are reflections of
  the filled symbols about midrapidity. The results derived from
  measurements of $E_T$ and $N_c$  
  flow~\cite{l877flow2} for nucleons and pions are shown by stars~(see text).
  The dotted line is added to aid
  inspection of the solid stars for antisymmetry about mid-rapidity. 
  Results from calculations using 2 versions of RQMD (1.08 in cascade mode,
  2.3 in mean field mode) are depicted as histograms.
    }
\end{figure}

\begin{figure}
\centerline{\psfig{figure=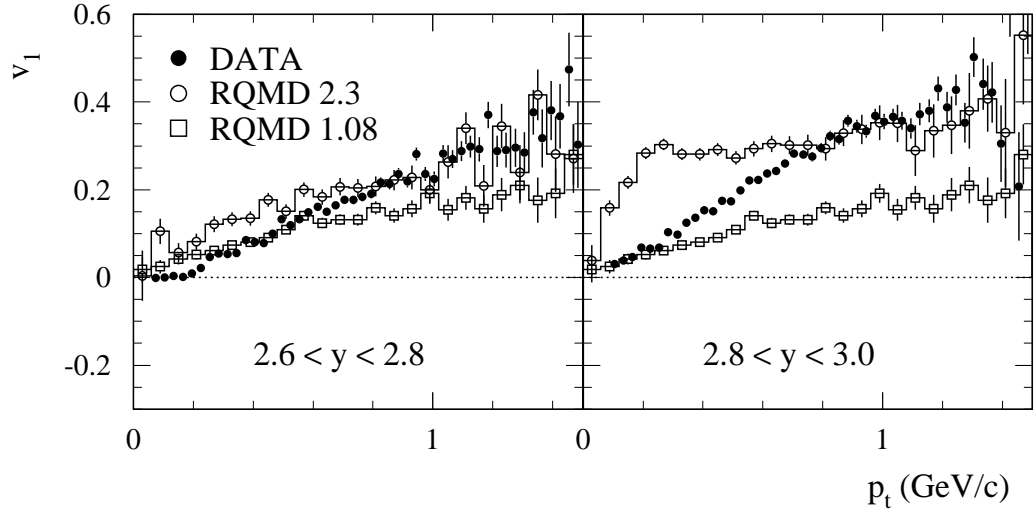,height=9.0cm}}
  \caption[]{
  Comparison of measured values of $v_1$ for protons in two rapidity
  bins and the centrality
  bin with $E_T$ = 200-230 GeV with those predicted by 2 versions of the
  event generator RQMD. 
    }
\end{figure}

\end{document}